\documentstyle[12pt,fleqn]{article}
\input{epsf}

\newcommand{\be}{\begin{equation}}
\newcommand{\ee}{\end{equation}} 
    \newcommand{\ha}{\frac{1}{2}}
  
\newcommand{\noi}{\noindent}
\newcommand{\ap}{anisotropy and polarization}
\newcommand{\ga}{\alpha}

\newcommand{\gc}{\gamma}
\newcommand{\gd}{\delta}

\newcommand{\gth}{\theta}

\newcommand{\gl}{\lambda}

\newcommand{\gn}{\nu}

\newcommand{\gs}{\sigma}

\newcommand{\gt}{\tau}

\newcommand{\gff}{\varphi}

\newcommand{\gw}{\omega}

\newcommand{\gD}{\Delta}

\newcommand{\gL}{\Lambda}

\newcommand{\gS}{\Sigma}

\newcommand{\gW}{\Omega}

\newcommand{\ti}{\widetilde}

\newcommand{\ra}{\rightarrow}

\newcommand{\ov}{\overline}

\newcommand{\lan}{\langle}
\newcommand{\ran}{\rangle}

\begin{document}
{\hbox to\hsize{}
\begin{center}
{\Large \bf {The Cosmological Gene Project: cluster analysis of the 
atmospheric 
fluctuations on arcmin-scale imaging of the Cosmic Microwave 
background}}\\
\bigskip
{\bf H.E.J\o rgensen}\\

{\it Astronomical Observatory, Juliane Maries Vej 30, DK-2100 Copenhagen,
Denmark}\\
{\bf E.V.Kotok}\\
{\it Theoretical Astrophysics Center, Juliane Maries Vej 30, DK-2100
Copenhagen, {\O} Denmark}\\
{\bf I.P. Naselsky, P.D.Naselsky,}
\footnote{Also: Theoretical Astrophysics Center,
Juliane Maries Vej 30, DK-2100 Copenhagen;
NORDITA, Blegdamsvej 17, DK-2100 Copenhagen, Denmark.}\\
{\it Rostov State University, Zorge 5, 344090 Rostov-Don, Russia}\\
{\bf I.D.Novikov}
\footnote{Also: Astro-Space Center of Lebedev Physical Institute,
Profsoyuznaya 84/32, Moscow, Russia; Astronomical Observatory,
Juliane Maries Vej 30, DK-2100 Copenhagen, Denmark;
NORDITA, Blegdamsvej 17, DK-2100 Copenhagen, Denmark.}\\
{\it {Theoretical Astrophysics Center,
 Juliane Maries Vej 30, DK-2100 Copenhagen, Denmark}}\\
{\bf Yu.Parijskij, P.Tcibulev}\\
{\it Special Astrophys. Obser., Nizhnij Arkhyz, Karachaj-Cherkess 
Republic, 357147 Russia.}\\
{\bf E.V.Vasil'ev}\\
{\it Rostov State University, Zorge 5, 344090 Rostov-Don, Russia}\\
\end{center}
\date{\today}

\begin{abstract}
We discuss some aspects of the Cosmological Gene Project started at the 
Special Astrophysical Observatory (Russia) in 1999. The goal of the 
project is to measure the anisotropy and polarization  of the Cosmic 
Microwave Background (CMB) and investigation of atmospheric fluctuations 
and foreground on arcmin-scales using the  radio-telescope RATAN-600.
We develop the cluster analysis of one-dimensional random fields 
for the application to the RATAN-600 scans. We analyze the specific 
properties of peak clusterisation in the RATAN-600 scans which to 
separate the primordial CMB signal from noise.
\end{abstract}

\section{Introduction}
The Cosmological Gene Project started at  the Special Astrophysical Observatory
(SAO RAN, Russia) in 1999 and is devoted to the measurements 
of \ap \ of the Cosmic Microwave Background (CMB) with the highest possible
angular and multipole resolution (up to $l\le10^5$)\footnote{General
information about CG-project is available on the web-site www.sao.ru}.
This project is carried out by the radio telescope RATAN-600. 
This is a reflector type
instrument, world largest in size and specially designed for high sensitivity
and high angular resolution multi-wavelengths measurements in the
frequency interval $\nu=1-30$GHz being the optimum for ground based telescopes.
For all ground based CMB \ap \  projects it is important to know the level of 
the atmospheric emission from water molecules since water vapor 
in the troposphere close to the condensation point [1,2] is the 
strongest source of brightness temperature variation.

Investigation of the atmospheric emission (AE) at frequency  $\nu=1-30$GHz 
has been done on RATAN-600 during more than 20 years.
The detection level of AE has approached an antenna temperature of a few
$m$K while the CMB signal could be close to a few $\mu$K. Thus, in order to
reach the level of primordial CMB signal the efficiency of AE filtration has
to be extremely high $\succeq 10^3$. It can be obtained by accumulation
of observational data at different frequencies and by long term
averaging.

Besides that, due to the very large diameter of the antenna ($D\simeq 600$m) 
the atmospheric noise lies in a so called ``near-field zone'' of the instrument 
and are smoothed by the aperture averaging effect. In addition the dual-beam 
and multi-frequency filtration of the AE signal are especially important for 
cleaning of the observational records.

The critical point of the AE filtration problem is the estimation of the 
damping factor of the AE variance. The two following points play an important 
role for resolving this problem. Firstly, after the CMB experiments 
already carried out (see for review [3]) 
we can predict the general character of the CMB 
anisotropy power spectrum $C(l)$ at the high multipole range $l\gg10^2$. 
For the most interesting cosmological model (SCDM, $\gL$CDM, OpenCDM, 
etc.) the important property of the spectrum is the exponential 
decrease at $l\succeq l_d$ where $l_d \approx 10^3$ corresponds  the 
manifestation of the diffusion damping during recombination.
This means that at small angular scale
$\gth\le1 $ arcmin ($l\gg l_{d}$) the observational 
signal only contains the AE noise and foreground noise (mainly unresolved 
point sources  PS).

Following [4] we can expect that the power spectra at $l\gg l_{d}$
are: $C_{PS}(l)\propto const$ and $C_{AE}\propto l^{-3}$ for unresolved 
point sources and atmospheric emission respectively. 
This means that the extra resolution of the RATAN-600 antenna beam 
(up to $l\sim 10^4 -10^5$) allows us to determine the structure of the 
noise signal at the extremely small angular scale. The next point is 
connected with the cluster analysis and peak statistics of the 
CMB+AE+foreground signal in the RATAN-600 scans. We would like to point 
out that in every 24$^h$ record of the radio sky we can ``see'' 
$\sim l_{d}\simeq 10^3$ maxima and minima connected with the CMB signal 
and $\sim 10^{4}-10^5$ peaks corresponding to all types of noises 
(radiometer, AE, PS and so on). In another words, in every 
RATAN-600 scan we have a wide represented peak statistics due to the 
CMB signal and the different types of noise as well.

The second point is related to the statistical properties of the 
observational signal. One of the most important prediction of the 
inflation paradigm is the Gaussian character of the CMB anisotropy 
distribution on the sky. For the AE noise and foreground noise Gaussian
fluctuations could be ``a lucky chance'' only. Thus, in the general case being
the combination of the primordial CMB signal and different types of noise 
in RATAN-600, scans can be non-Gaussian due to the statistical properties of 
the noise.

P. Naselsky and D. Novikov [5] 
and D. Novikov and H. J\o rgensen [6]
pointed out that in this case a cluster analysis of the signal is 
very effective to separate noise from the signal. The goal of this work is to analyze how the cluster analysis 
technique can be applied to separate the primordial CMB signal from noise 
(AE and PS) for the particular case of observations with RATAN-600.

\section{Cluster analysis of the one-dimensional random fields}

As we mentioned above, the two-dimensional random CMB+AE+foreground signal
on the sky is transformed by the RATAN-600 antenna-beam into series of
the one-dimensional realization with very high angular resolution. In
Fig.1a we plot the result of the weighted average of the 57 records of
the AE noise which was obtained at wave-length $\gl=1.38$cm at declination
$\gd=41^o$ and angular length of scan $L=11.3^o$ during the 1998
observational period.
\begin{figure}[h]
\vspace{0.01cm}\hspace{-0.1cm}\epsfxsize=7.5cm 
\epsfbox{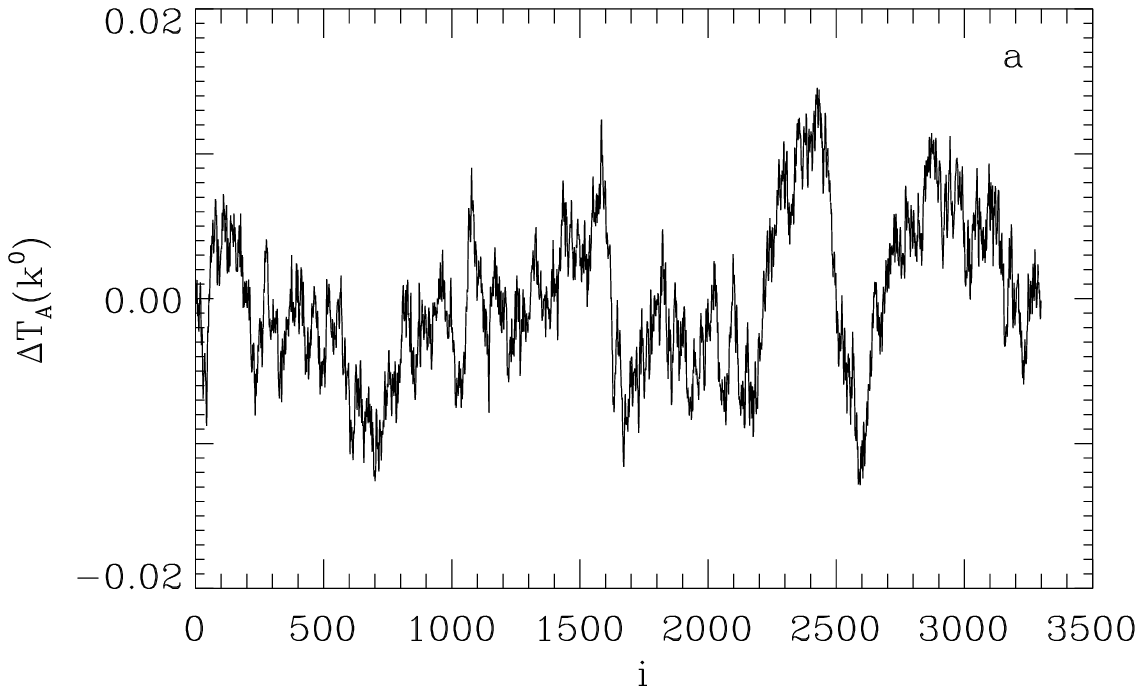}
\hspace{0.0cm} \epsfxsize=7.5cm 
\epsfbox{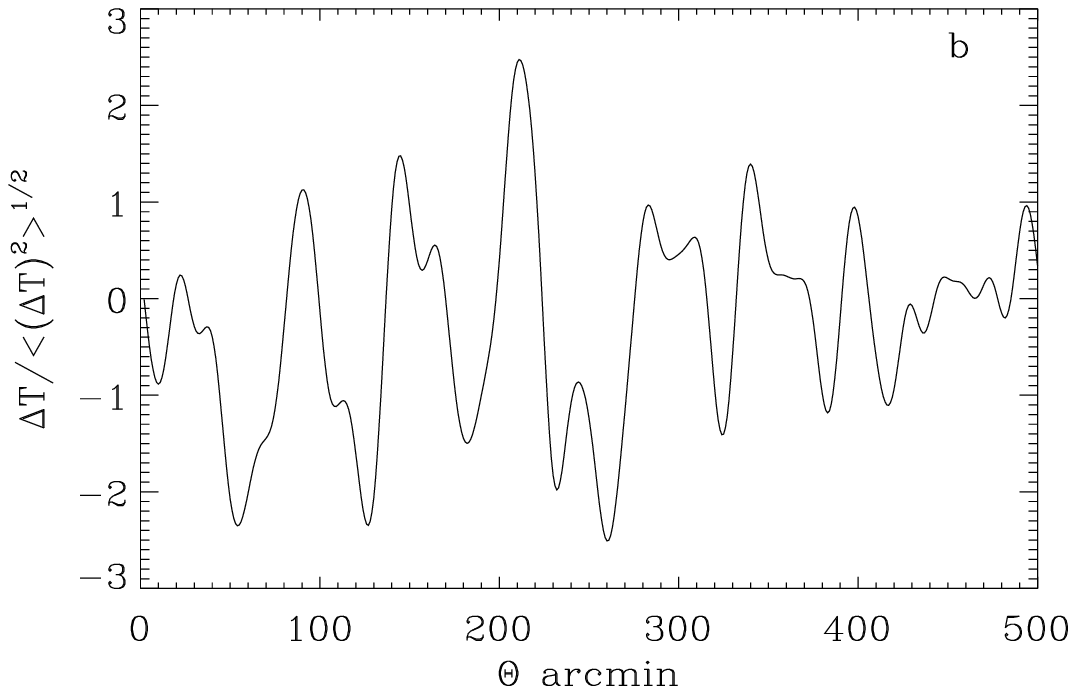} 
     \caption{ {\it (a) Scan of weighted average of the 57 records of 
AE signals. On the $x$-axis $i$ indicates the step number on the scan with 
the length $L=11.3^o$ and with the step size $\gD x=12$arcsec. 
(b) One-dimensional realization of the primordial CMB signal.}}
\end{figure}

Fig. 1b shows one theoretical realization realization of
the primordial CMB signal,  normalized to the variance, which we simulated 
for the standard cold dark matter model (SCDM). As one can see from
Fig1a,b
both signals, CMB and AE, appear approximately uniform and have a lot of
maxima and minima. Let us normalize pure CMB fluctuations and AE noise to their
variance: $\nu=\gD T/\sqrt{\lan\gD T^2\ran}$ and consider  maxima and
minima above some threshold $\nu_t$. Following [5,6] 
we introduce the definition of a one-dimensional cluster of maxima 
as the continuous part of the curve with $\nu(t)>\nu_t$ inside the interval
$t\in(t_1,t_2)$ where $t_1$ is one of the roots of the equation 
$\nu(t_{1,2})=\nu_t$ and $t_2$ is the closest root to $t_1$, $t_2>t_1$.
The length $k$ of the cluster is defined as the number of peaks with height
$\nu_{peak}>\nu_t$ in the cluster.
If the value of $\nu_t$ is high ($\nu_t\gg 1$) then only high maxima of
the random field are present above the threshold and the typical
length of a cluster is $k=1$.
The reduction of the threshold level $\nu_t$ down to $\nu_t\ra0$ or
$\nu_t<0$ leads to the appearance of big clusters when maxima of smaller clusters
begin to connect together and generate a new cluster. The
investigation of the cluster statistics could have a great potential for
probing the nature of CMB and AE signals on the scans. For this purpose,
following [5,6],
we introduce the number of clusters $N_k(\nu_t)$ of the
length $k$ and the total number of clusters $N(\nu_t)$ which are presented
in the anisotropy scans for an appropriate threshold $\nu_t$.
\be
N(\nu_t)=\gS^{\infty}_{k=1}N_k(\nu_t)
\label{1}
\ee
and the mean length of a cluster at cross level $\nu_t$:
\be
\lan k(\nu_t)\ran=
\frac{\gS^{\infty}_{k=1}kN_k(\nu_t)}{\gS^{\infty}_{k=1}N_k(\nu_t)}.
\label{2}
\ee
Note that the definitions Eq.(\ref{1}) and Eq.(\ref{2}) do not depend on the
nature of the initial random signal or its composition.

Let us turn back to the data of AE (see Fig.1a) and take into account that the
step on the x-axis is only $\gD x=12$arcsec.
Such a high angular resolution is wittingly in excess for the analysis of the 
real data and allows  us to use an additional procedure of filtration
of noise by smoothing the data. Let us consider the following smoothed array of 
data $\Psi_i$:
\be
\Psi_i=\frac{\gS_j f_i K_{ij}(d)}{\gS_j  K_{ij}(d)},
\label{3}
\ee
where $f_i$ is the initial array of data of Fig.1a, $K_{ij}(d)$ is the filter 
(see below) with the characteristic scale of smoothing $d$.

\begin{figure}[h]
\vspace{0.01cm}\hspace{2cm}\epsfxsize=8cm 
\epsfbox{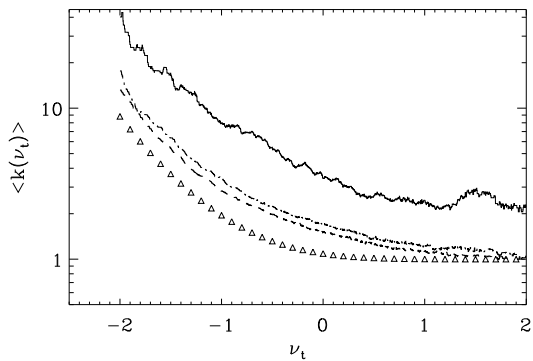}
     \caption{ {\it The mean length of a cluster. From top to bottom:
initial AE signal from Fig.1a, smoothed AE signals ($d=5$ and $d=10$ 
correspondingly), pure CMB signal represented by triangles for the SCDM.}}
\end{figure}

We use below (as one of the possible choices) a Gaussian filter
$K_{ij}(d)=\exp(\frac{(i-j)^2}{2d^2})$ for different values of parameter
$d$: $d=5$,  $d=10$ and  $d=100$. After such a filtration we can easily
simulate the rate of clusterisation for the initial and smoothed data.
The result of this simulation is represented on Fig.2. As one can see
from this figure, while the scale of smoothing $d$ increases from $d=5$
(which corresponds to the characteristic scale $\ga\simeq 1$ arcmin) up to
$d=100$ ($\ga\simeq 20$ arcmin) the mean length of the clusters 
decreases monotonically for a fixed level of the threshold. One can compare 
this result with the rate of clusterisation (mean length of clusters) of the 
pure CMB signal (see Fig.2) which is found systematically less than the non 
smoothed AE data. This tendency become even more pronounced if we use the 
procedure of filtration of the pure CMB signal with $d=5-10$.

Thus, as one can see from the above result, if we have the same scale of
smoothing for the AE signal and the pure CMB signal then the rate of
clusterisation of AE noise  is always larger than the rate of
clusterisation of the pure CMB signal.

In the following we will discuss the theoretical rigor 
of this phenomenon more precisely.
The critical point of our analysis is the hypothesis that the primordial
CMB anisotropy distribution on the sky is Gaussian. 
Observations by RATAN-600 give one-dimensional scans of $\gD T$. Let us 
consider a theoretical realization of a one-dimensional slice with $\L\ll2\pi$ 
of the two-dimensional $\gD T$-field with power spectrum $C_T(l_{ij})$. 
$\gD T(x)$ for this one-dimensional slice can be written as follows [5,6]:
\be
\gD T(x)=\sum_{i,j}a_{ij}\sqrt{C_T(l_{ij})}\cos\left(\frac{2\pi}{L}ix+
\gff_{ij}\right)W_{ij},
\label{4}
\ee
where $a_{ij}$ are independent Gaussian numbers, $C_T(l_{ij})$ is the power
spectrum of fluctuations, $\gff_{ij}$  are random phases homogeneously
distributed in the interval $(0,2\pi)$,
$l_{ij}=\frac{2\pi}{L}\sqrt{i^2+j^2}$, $W_{ij}$ is the RATAN-600 antenna-beam.
For the uniform Gaussian random process Eq.(\ref{4}) the two point
correlation function has the standard form:
\be
C_{obs}(\gt)=\lan\gD T(x)\gD T(x') \ran=
\ha\sum_{i,j}a^2_{ij}J_0(l_{ij}\gt)C(l_{ij})W^2_{ij},
\label{5}
\ee
where $\gt=|x-x'|$. After averaging over the ensemble the correlation
function and its variance are [6]:

\be
C(\gt)=\frac{\pi}{2}\int dl\ C(l)J_o(l\gt)W^2(l)
\label{6}
\ee
\[
D(\gt)=\ov{C^2_{obs}(\gt)}-[\ov{C_{obs}(\gt)}]^2=
\frac{\pi}{2}\int  dl\ C^2(l)J^2_o(l\gt)W^4(l).
\]
For investigation of the peak statistics we can introduce
the spectral parameters [7]:
\be
\gs^2_p=p ! 2^{2p}(-1)^p\frac{d^pC(\gt)}{d(\gw^2)}\left|_{\gt=0}\right.,
\label{7}
\ee
where $\gw=2\sin\frac{\gt}{2}$. As it is well known, the following parameters
and its combination are the most important for peak statistics:
\be
\begin{array}{lc}
\gs_0^2=\pi\int dl lC(l)W^2(l); \hspace{0.5cm}
\gs_1^2=\pi\int dl l^3C(l)W^2(l); \hspace{0.5cm}
\gs_2^2=\pi\int dl l^5C(l)W^2(l)\\
\gamma=\gs_1^2/(\gs_o\gs_2);\hspace{1.0cm} R_*=\gs_1/\gs_2.
\end{array}
\label{8}
\ee

D. Novikov and H. J\o rgensen [6] have shown that the value of parameter
$\gamma$ determines the topology of the $\gD T$ distribution on the
sky as well as the rate of peak clusterisation which decreases with $\gamma$. 
According  to [6]
\  the mean length of clusters at threshold $\nu_t$ is
\be
\lan k(\nu_t)\ran=[1-\ga(\nu_t,\gamma)]^{-1},
\label{9}
\ee
where $\ga(\gn_t\gamma)=\ti{N}^+_{min}(\nu_t)/\ti{N}^+_{max}(\nu_t)$ and
\be
\ti{N}^+_{max}=\frac{1}{4\pi}\frac{\gs_2}{\gs_1}
\left\{1-\Phi\left(\frac{\nu_t}{\sqrt{2(1-\gamma^2)}}\right)+
\gamma e^{-\frac{\nu^2_t}{2}}
\left[1+\Phi\left(\frac{\gamma\nu_t}{\sqrt{2(1-\gamma^2)}}\right)\right]\right\},
\label{10}
\ee

\be
\ti{N}^+_{min}=\frac{1}{4\pi}\frac{\gs_2}{\gs_1}
\left\{1+\Phi\left(\frac{\nu_t}{\sqrt{2(1-\gamma^2)}}\right)-
\gamma e^{-\frac{\nu^2_t}{2}}
\left[1-\Phi\left(\frac{\gamma\nu_t}{\sqrt{2(1-\gamma^2)}}\right)\right]\right\},.
\label{11}
\ee
where $\Phi(x)$ is the error function [8].
\ In the next sections we discuss general  properties of the parameter 
$\gamma$ for different 
types of CMB spectra, $C(l)$, corresponding to different cosmological 
models (SCDM, $\gL$CDM, etc). Different strategies for the
RATAN-600 observations will also be discussed.

\section{Statistical properties of the $\gD T$-signal in the RATAN-600 scans.}

In the framework of Cosmological Gene Project
several modes (or strategies) of observations are possible (see [9]).
\ Below we discuss only the SECTOR mode of radio sky observations where
several $\frac{1}{4}$ parts of the RATAN-600 antenna ring with separated 
secondary mirrors may be used. Besides that, the beam-switching strategy 
in the SECTOR mode is very important for both AE and point sources 
filtration and detection of the primordial CMB signal. 

Let us discuss one of the possible beam-switching  regimes. In this case 
two radiometers are turned to one secondary mirror and measure the 
difference of the $\gD T$-signals between themselves:
\be
\ti{\gD T}_{II}(x)=\gD T(x+d)-\gD T(x),
\label{12}
\ee
where $d$ is the angular distance between pointings of the radiometers.
We find that 
\be
\ti{\gD T}_{II}(x)\approx \frac{dT}{dx}|_{d=0}\ d
\label{13}
\ee
for the multipole numbers $ld\ll1$. If $d^{-1}$
corresponds to $l \sim 10^4-10^5$ then  Eq.(\ref{13}) is correct practically 
in all of the most important ranges of the observations for 
$l\le 10^4-10^5$.
Note, that due to the linear character of the  Eq.(\ref{13}),
the distribution of $\ti{\gD T}_{II}(x)$ is Gaussian. 
More over, the multipole spectrum of function $\gD T_{II}(x)$ is
connected with the multipole spectrum of the initial signal as 
\be
\ti{C}_{II}(l)=l^2d^2C(l).
\label{14}
\ee
Thus, as we can see, the analysis of the statistical properties of 
the first derivatives of the initial signal is practically equivalent 
to the analysis of the power spectrum $l^2C(l)$. In addition there are two
main points which are very important for the future investigations.
The first is the following. As we mentioned above, the high frequency 
noise, which is connected with the noise of a radiometer and other 
sources,
are usually present in the observational atmospheric emission scans
of RATAN-600  (see for example Fig.1a). These high frequency 
modulations can be extracted from the initial scan by smoothing over 
scales $\gth_S\simeq 5-15$arcsec $>d$ using a Gaussian filter for example:
\be
\gD T_{II}(x)=\frac{1}{\sqrt{2\pi}\gth_S}\int^{\infty}_{-\infty}
\ti{\gD T}_{II}(x')\exp\left[-\frac{(x-x')^2}{2{\gth_S}^2}\right]dx'.
\label{15}
\ee
The smoothing procedure in Eq.(\ref{15}) leads to the following 
transformation of the power spectrum Eq.(\ref{14}):
\be 
C_{II}(l)=l^2d^2C(l)e^{-l^2\gth^2_s}.
\label{16}
\ee
The second point is connected with general properties of the primordial 
spectrum $C(l)$ for the most popular cosmological models at high multipoles
$l\ge10^2$.
As we mentioned in the introduction the spectrum has an exponential decrease 
due to damping at $l\succeq l_d\approx10^3$: $C(l)\propto exp{(-l^2R^2_d)}$ 
where $R_d\sim l_d^{-1}$.

\begin{figure}[h]
\vspace{0.01cm}\hspace{3cm}\epsfxsize=7.5cm 
\epsfbox{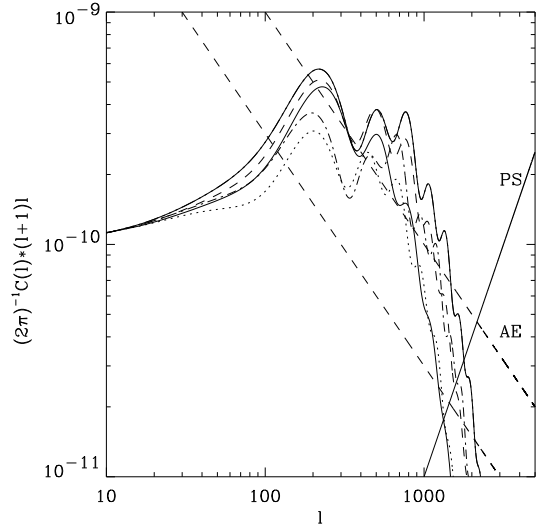}
     \caption{ {\it Spectra of primordial CMB signal for different 
cosmological models (see Table 1). Straight dashed lines represent the
spectrum of AE and the straight solid line represents spectrum for 
unresolved point sources (PS).}}
\end{figure}

Besides that, for the Harrison-Zeldovich initial power spectrum of
adiabatic perturbations the typical trend of the multipole spectrum of CMB
is $C(l)\propto l^{-2}$ at $l_d>l\gg 1$ with Sakharov's modulations. 
Moreover, we expect that for the AE
fluctuations the spectrum of the noise is close to a power law spectrum as well.
Similarly we get for the different types of the foreground,
synchrotron emission, dust and free-free emission, $C(l)\sim l^{-3}$ 
[4] and $C(l)\sim l^0$ for the point sources.

Thus, for the future investigation we adopt the following form of 
the considered power spectrum $C_{II}(l)$:
\[
C_{II\  CMB}(l)\propto d^2e^{-l^2(R_d^2+\gth^2_S)}l^q {\hspace{2.0cm}}
{\normalsize {for\  the\  primordial\  CMB\  signal}}
\]
\be
C_{II\  PS}(l)\propto d^2l^2e^{-l^2\gth^2_S} {\hspace{3.2cm}}
{\normalsize {for\  PS\  noise}}
\ee
\label{17}
\[
C_{II\  atm}(l)\propto d^2 l^2(1+l^2a^2)^{-m}e^{-l^2\gth^2_S} 
{\hspace{0.9cm}}
{\normalsize {for\  AE\  noise}}
\]
where $\Phi(l)=l^qe^{-l^2R^2_d}$ describes Sakharov's modulations in the CMB spectra,
$a^{-1}$ is the typical scale in $l$-space for the power part of the AE 
spectrum; $m$ and $q$ are spectral indices.
\begin{figure}[h]
\vspace{0.01cm}\hspace{1cm}\epsfxsize=7.5cm 
\hspace{0.8cm} \epsfxsize=6cm \epsfbox{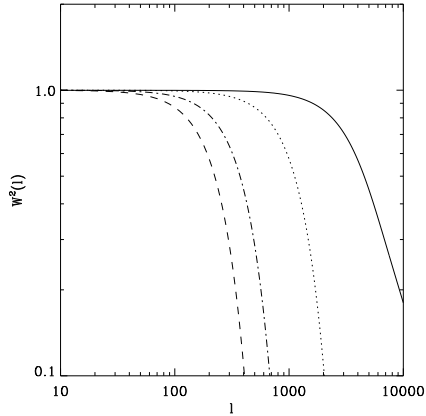}
    \caption{ {\it Left panel represents the beam profile of the 
RATAN-600 antenna for the SECTOR mode at $\gl=1$cm. The right panel 
$W^2(l)$: solid line for RATAN-600 scans, dotted line for Planck HFI,
dashed-dotted line for MAP and dashed line for  PLANCK LFI.}} 
\end{figure}

The last point is connected with the shape  of the antenna beam.
For example, in the MAP and PLANCK satellite missions the Gaussian antenna 
beams correspond to $W^2(l)\propto \exp(-l^2\gth^2_A)$,
$\gth_A \simeq 7.45\times 10^{-3}\left(\frac{\gth_{FWHM}}{1^o}\right)$ 
[10],
\ where $\gth_A = 2.23\times 10^{-4}$ for the MAP project 
$(\gth_{FWHM}=0.3^o$) and $\gth_A = 7.45\times 10^{-4}$ for the HFI 
detector in PLANCK ($(\gth_{FWHM}=0.1^o$).
The multipole character of the RATAN-600 antenna beam is more 
complicated than antenna beams of MAP and PLANCK. 
For $W(x,y)\ge 0.1$ we use the following 
expression [11]:
\be
W(x,y) =\exp \left\{-(\gamma 'y)^2-\ha \ln[1+(\ga y)^2]
-\frac{(\gd  x)^2}{1+(\ga y)^2}\right\}, 
\label{18}
\ee
where $\ga =\frac{2\sqrt{3}}{FWHM(y)}$, 
$\gd =\frac{2\sqrt{\ln2}}{FWHM(x)}$, 
$\gc '=\frac{2\sqrt{\ln 2}}{FWHM(E)}$, and $FWHM$ is the standard 
definition of the ``full-width-half-maximum''
in the $x$ and $y$ directions and energy respectively. This description is 
applicable for the central part of the antenna beam and the parameters 
$\ga ,\gd $ and $\gc '$ depend on the frequency of the observations. 
For example, at 
$\nu =30$GHz the values of $FWHM$ are the following:
$FWHM(x) =5.1$ arcsec, $FWHM(y) =37.7$ arcsec and $FWHM(E) =717$ arcsec.

We determine the Fourier transformation of  $W(x,y)$:
\be
W_{ij}=(1+k^2_{\gamma}\mu^2)^{-1/2}\exp\left[-\frac{k^2_i}{4\gd^2}-
\frac{k^2_j}{4(\gamma^2+\frac{k_i^2\ga^2}{4\gd^2})}\right],
\label{19}
\ee
where $k_i=\frac{2\pi\ i}{L}$, $k_j=\frac{2\pi\ j}{L}$, 
$\mu=\ga/2\gamma\gd$ and substitute $W_{ij}$ from Eq.(\ref{19}) in 
Eq.(\ref{5}). After the transformation $l=\frac{2\pi}{L}\sqrt{i^2+j^2}$ 
and integration over angle $\gff=\arctan j/ i$ we find the dependence 
$W^2(l)$ on $l$ as follows
\be
W^2(l)=\frac{2}{\pi}\int_0^1\frac{dx \ \exp\left\{-P(x)\right\}}
{\sqrt{1-x^2}(1+m^2(l)x^2)}
\label{20}
\ee
where $P(x)=d^2(l)x^2+\frac{c^2(l)(1-x^2)}{1+m^2(l)x^2)}$, \ 
$d(l)=\frac{l}{\sqrt{2}\gd}$,\  $c(l)=\frac{l}{\sqrt{2}\gamma}$, 
$m(l)=l\ \mu$.

In Fig. 4 we have plotted $W^2(l)$ from Eq.(\ref{20}). 
As one can show from this figure we can apply the following approximation 
for  $W^2(l)$ practically up to $l\sim 10^4$:
\be
W^2_{RATAN}(l)\simeq \left(1+\frac{l^2}{l_0^2}\right)^{-1},
\label{21}
\ee
where $l_0=4.5\times10^3$.

Taking into account notations mentioned above (see Eq.(\ref{7}) and 
Eq.(\ref{8})), we determine spectral parameters $\gs^2_p$ for $C_{II}(l)$ 
from Eq.(\ref{17}) and definition 
Eq.(\ref{9}):
\be
\gs^2_{p,CMB}=
\frac{\pi\ d^2\ A}{2(l_0^2)^{p+1+q/2}}\Gamma\left(p+1+\frac{q}{2}\right),
\label{22}
\ee
\be
\gs^2_{p,PS}=
\frac{\pi\ d^2\ B}{2\gth_S^{2(p+2)}}\Gamma\left(p+2\right),
\label{23}
\ee
\be
\gs^2_{p,AE}=
\frac{\pi\ d^2\ C}{2a^2\gth_S^{2p+1}}\Gamma\left(\ha+p\right),
\label{24}
\ee
where $\Gamma(x)$ is the $gamma$-function. For the AE spectrum in 
Eq.(\ref{17}) we took $m=3/2$, and $A$, $B$, $C$ are amplitudes of 
the CMB, PS and AE power spectra.

Let us discuss some exact values of the parameter $\gamma$ for the CMB, 
PS and AE signals. For $\gs^2_p$ 
from Eq.(\ref{22})-Eq.(\ref{24}) all $\gamma$ parameters (Eq.(\ref{8})) have 
extremely simple expressions:
\be
\gamma_{CMB}=\sqrt{\frac{q+2}{q+4}};\hspace{1.0cm}
\gamma_{PS}=\sqrt{\frac{2}{3}}=0.82;\hspace{1.0cm}
\gamma_{AE}=\frac{1}{\sqrt{3}}=0.58.
\label{25}
\ee
For realistic values of $q\simeq 0-2$ the corresponding values of 
$\gamma$ are $\gamma_{CMB} = 0.77$ at $q=1$ and 
$\gamma_{CMB}= \sqrt{\frac{2}{3}}=0.82$ at $q=2$.
As one can see from Eq.(\ref{25}) we have for the AE noise, $\gamma_{AE}\simeq0.58$ 
which is less than $\gamma_{CMB}$ and $\gamma_{PS}$. Therefore the rate 
of AE noise clusterisation is larger than the rate of the CMB 
and PS clusterisation referring to [6].

We compared our analytical predicted parameters
$\gamma$ with the data of numerical simulations of this parameter 
from real theoretical CMB spectrum $C_{II\ CMB}(l)$ using the CMBFAST code 
["""].
We used  the  results only of this numerical simulation and $W(l)$ for
the RATAN-600 antenna 
(see Fig.4b). The results are presented in
Table 1.
As one can see from this table, the values of the parameter $\gamma$ for all
investigated SCDM models are close to the value of 
$\ov{\gamma}_{CMB}\simeq 0.79$ and differing less
than 4\%. This means that the value of the parameter 
$\ov{\gamma}_{CMB}$ corresponds to $q\simeq 1-2$. For the $\gL$CDM model 
(see Table 1) a more preferable value of parameter $q$ is $q\simeq 0$.

Thus, as the result of our analytical investigation of the CMB, AE and PS 
spectra we have the following important conclusions. For different CMB 
power spectra the values of parameter $\gamma_{CMB}$ and the rate 
of clusterisation practically coincide. The AE noises due to more sloping
character of initial spectrum is clusterized more strongly than the primordial
CMB signal and  PS noise.

\section{Turning points of the peak clusterisation.}
As we have shown in the previous section,  the spectral parameters $\gs^2_p$ 
for different types of signals  characterize  the topology of
Gaussian maps corresponding to the different types of noise. In practice 
we have a combination of the primordial CMB signal, AE noise, 
foreground and PS. Therefore the topology of the map and the 
structure of a single scan are more complicated than topology of ``pure'' 
CMB, AE, PS and other signals. In this case the rate of peak clusterisation at 
different threshold levels $\nu_t$ is described in terms of the parameter 
$\gamma_{tot}$ if we suppose that all types of noise are Gaussian. Thus 
for mixed CMB+AE+PS+foreground sources the topology of a map can be 
described in terms of  $\gamma_{tot}$:
\be
\gamma_{tot}=\sum_{i=1}^N(\gs_1^2(i))
{\left\{\sum_{i=1}^N\gs_0^2(i)\sum_{i=1}^N\gs_2^2(i)\right\}}^{-1/2},
\label{26}
\ee
where $\gs^2_p(i)$ are given by Eq.(22)-(24),
$i$ is a number of the  type of noise and CMB signal,
$N$ is a total number of different types of noise and the CMB signal
\footnote{For the most important types of foreground such as synchrotron 
emission, dust and free-free emission  we have a power spectrum 
$C(l)\sim l^{-3}$. We 
denote a sum of this type of noises as $\gs^2_{AE}$, if the variance 
of the AE signal dominates. It is natural that methods of filtration for 
different types of noise can be different.}.\\
Let us introduce new functions:
\be
X=\frac{\gs^2_0(PS)}{\gs^2_0{(CMB)}}; \hspace{2cm}
Y=\frac{\gs^2_0(AE)}{\gs^2_0{(CMB)}}.
\label{27}
\ee
Using Eq.(26), (27) and (22)-(24) we have the following expression for 
the parameter
\be
\gamma_{tot}(X,Y,\xi)=\frac{1}{\sqrt{2}}\frac{\left[1+\xi(2X+\ha Y)\right]}
{\left[(1+X+Y)(1+\xi(3X+3/8Y)\right]^{1/2}},
\label{28}
\ee
where $\xi=1+\frac{R^2_d}{\gth^2_S}$ and we adopt $q=0$ for CMB signal.

Note that the parameter $\xi$  describes the dependency of the function 
$\gamma_{tot}$ of the angular resolution of the antenna beam and can be 
$\approx$1 if $\gth_S\gg R_d$ (for example for MAP project) and $\xi\gg1$ 
if $\gth_S\ll R_d$ (for the  RATAN-600 antenna beam). Thus, the angular 
sensitivity of the antenna plays a 
critical role in our cluster analysis technique. 


\begin{figure}[h]
\vspace{0.01cm}\hspace{-0.1cm}\epsfxsize=8cm 
\epsfbox{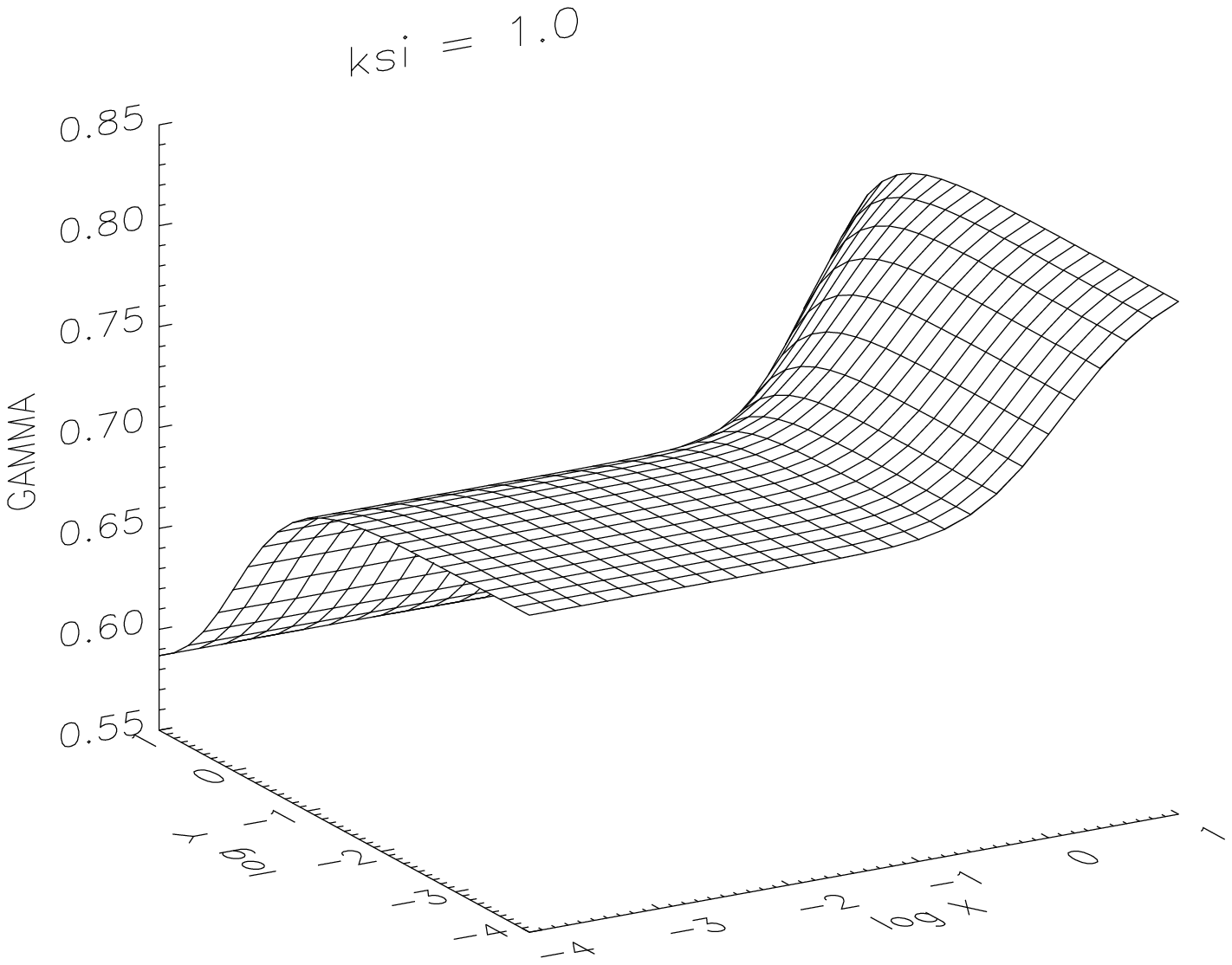}
\hspace{0.0cm} \epsfxsize=7.5cm 
\epsfbox{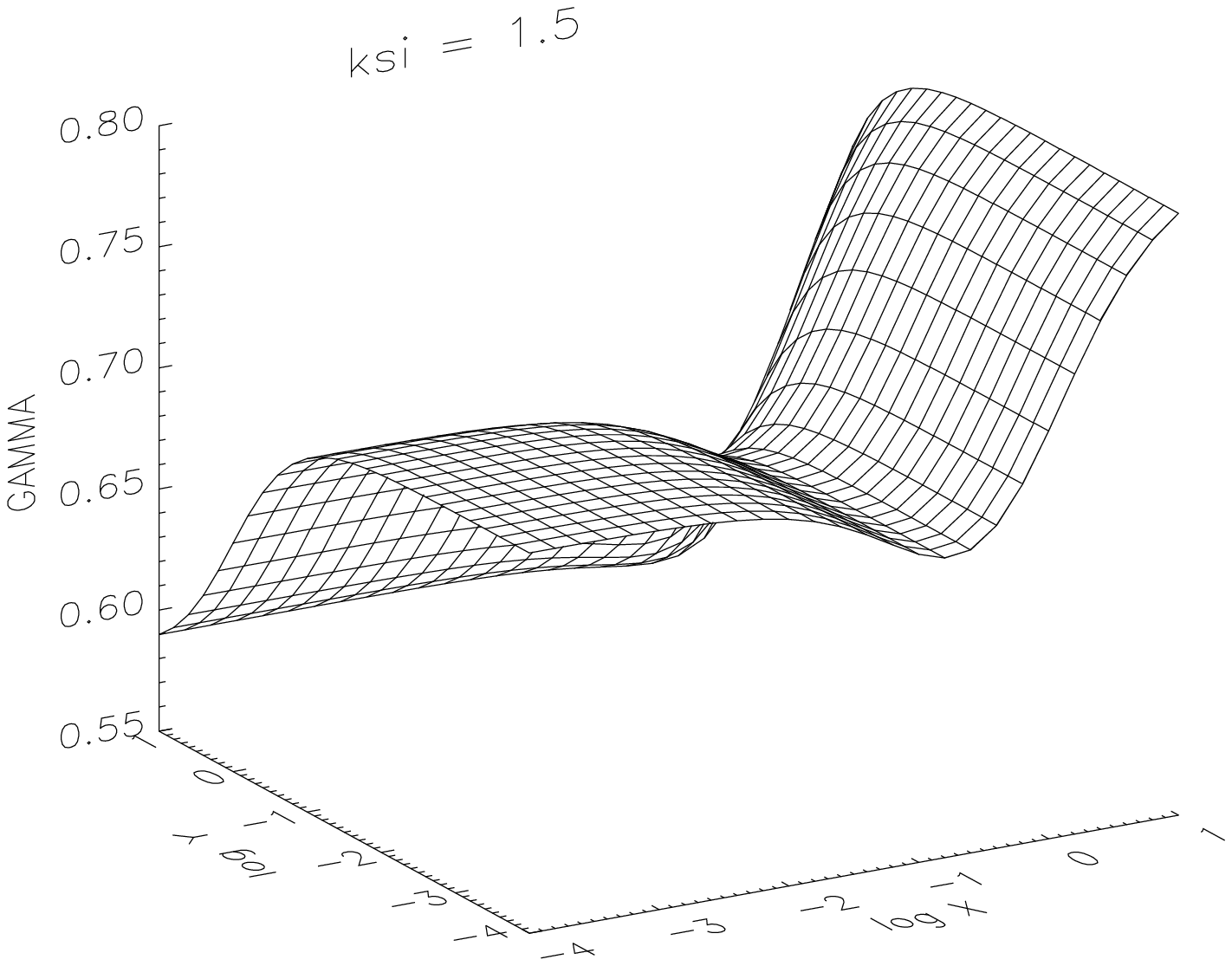} 
\vspace{0.01cm}\hspace{-0.1cm}\epsfxsize=8cm 
\epsfbox{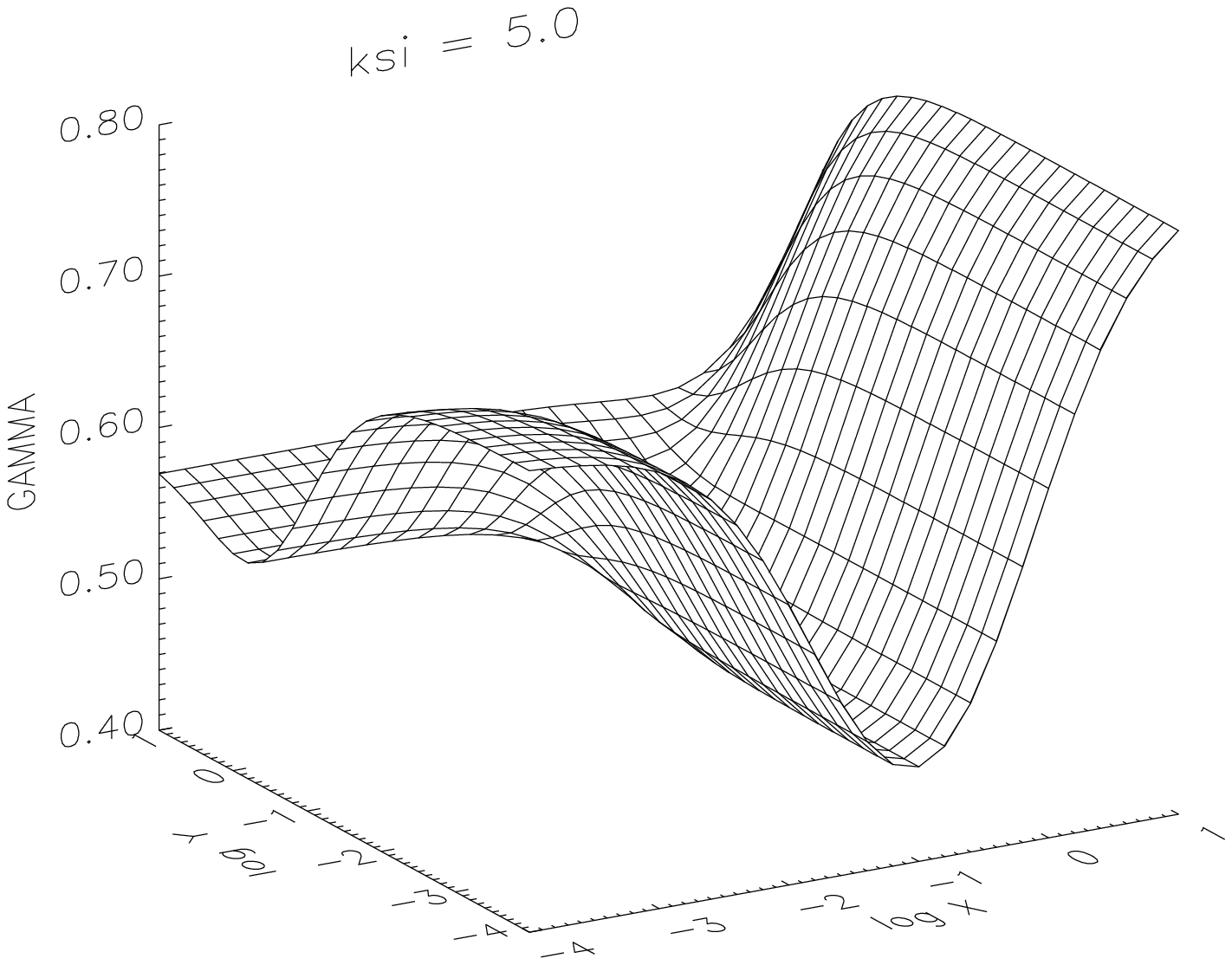}
\hspace{-0.1cm} \epsfxsize=7.5cm 
\epsfbox{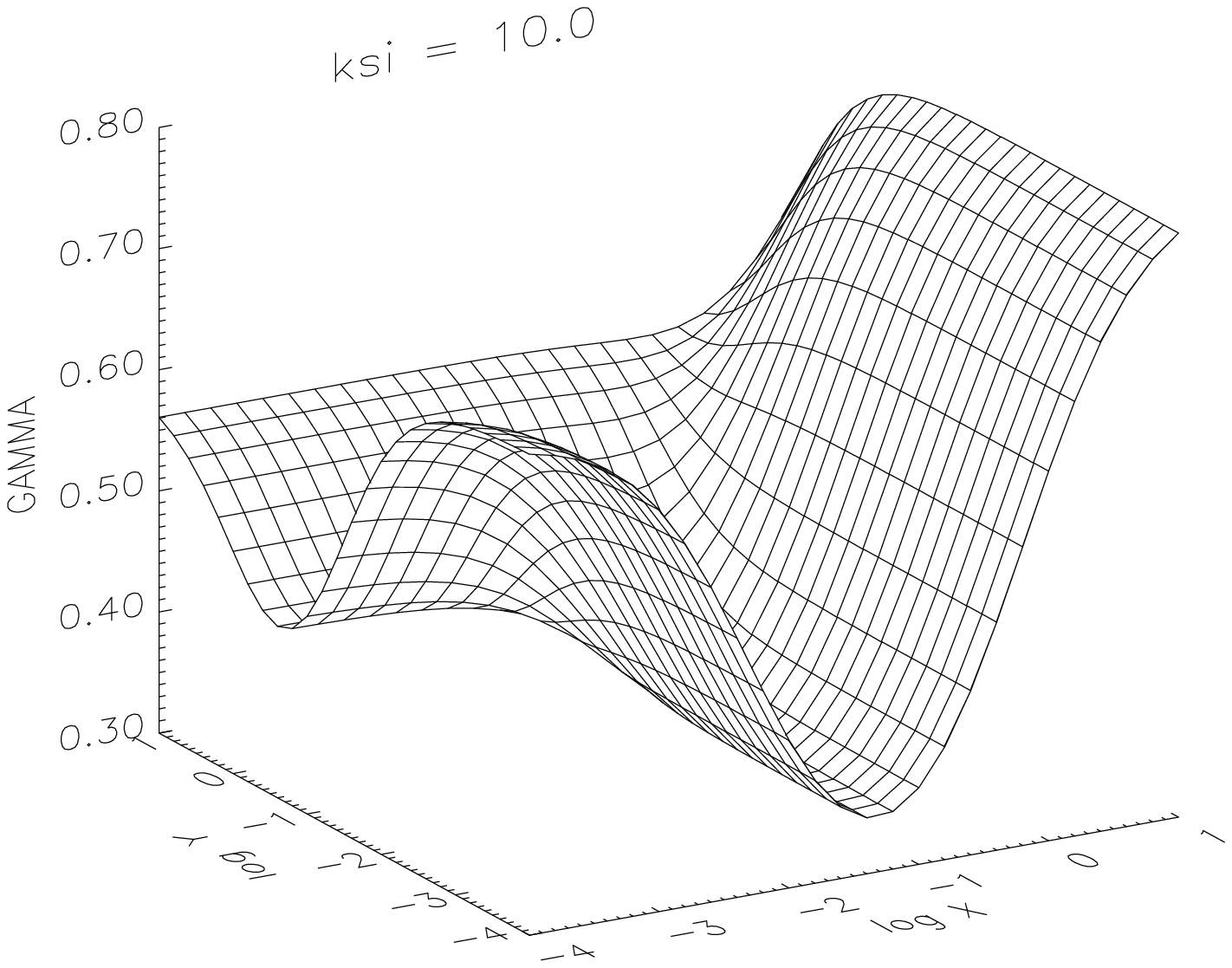} 
     \caption{ {\it $\gamma(X,Y,\xi)$-surfaces. Top left for $\xi=1$,
top right for $\xi=1.5$, bottom left for for $\xi=5$ and bottom right 
for $\xi=10$.}}
\end{figure}

In Fig. 5 we have plotted the function $\gamma_{tot}(X,Y,\xi)$ for different 
values of $\xi=1,1.5,5,10$. 
As seen from this figure, the shape of  $\gamma_{tot}(X,Y,\xi)$ 
surfaces at different values 
of parameter $\xi$ is very complicated. In Fig. 6 we represent 
two types of  cross sections of $\gamma_{tot}(X,Y,\xi)$ surfaces.
Panel a-c shows $\gamma_{tot}(X,Y_n,\xi)$ as function of X for 
$Y_n=10^{-3}- 1$ and $\xi=1,5,10$.
Panel d-f shows $\gamma_{tot}(X_n,Y,\xi)$ as function of Y for 
$X_n=10^{-3}- 1$ and $\xi=1,5,10$.

Thus the surface  $\gamma_{tot}(X,Y,\xi)$ describes the dependence of the 
main parameter of peak clusterisation based on different procedures of AE and 
foreground filtration. For the AE noise filtration one of the most important 
methods is the averaging of observational scans over ensembles.  
Due to the averaging procedure the variance of AE noise decreases in time as 
$t^{-1}$.  Increasing the observing 
time leads to a decrease of the parameter $Y$ (keeping $X=const$).
Using Fig. 5 we can estimate the trajectory $X(Y)$ which 
corresponds to any filtration procedure.

\begin{figure}[h]
\hspace{-0.01cm}\vspace{-0.01cm}
\epsfxsize=5cm 
\epsfbox{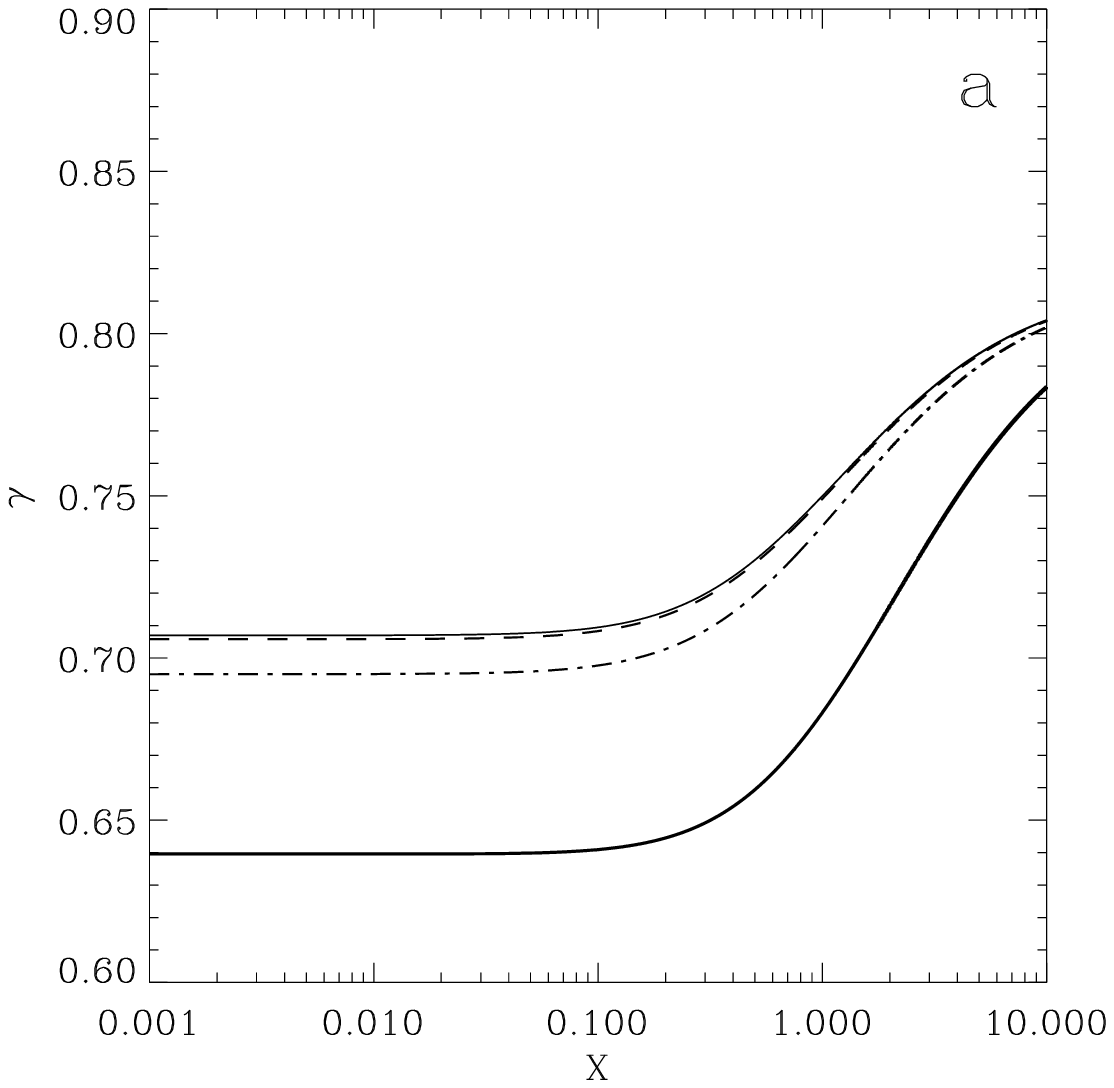}
\hspace{0.001cm} 
\epsfxsize=5cm 
\epsfbox{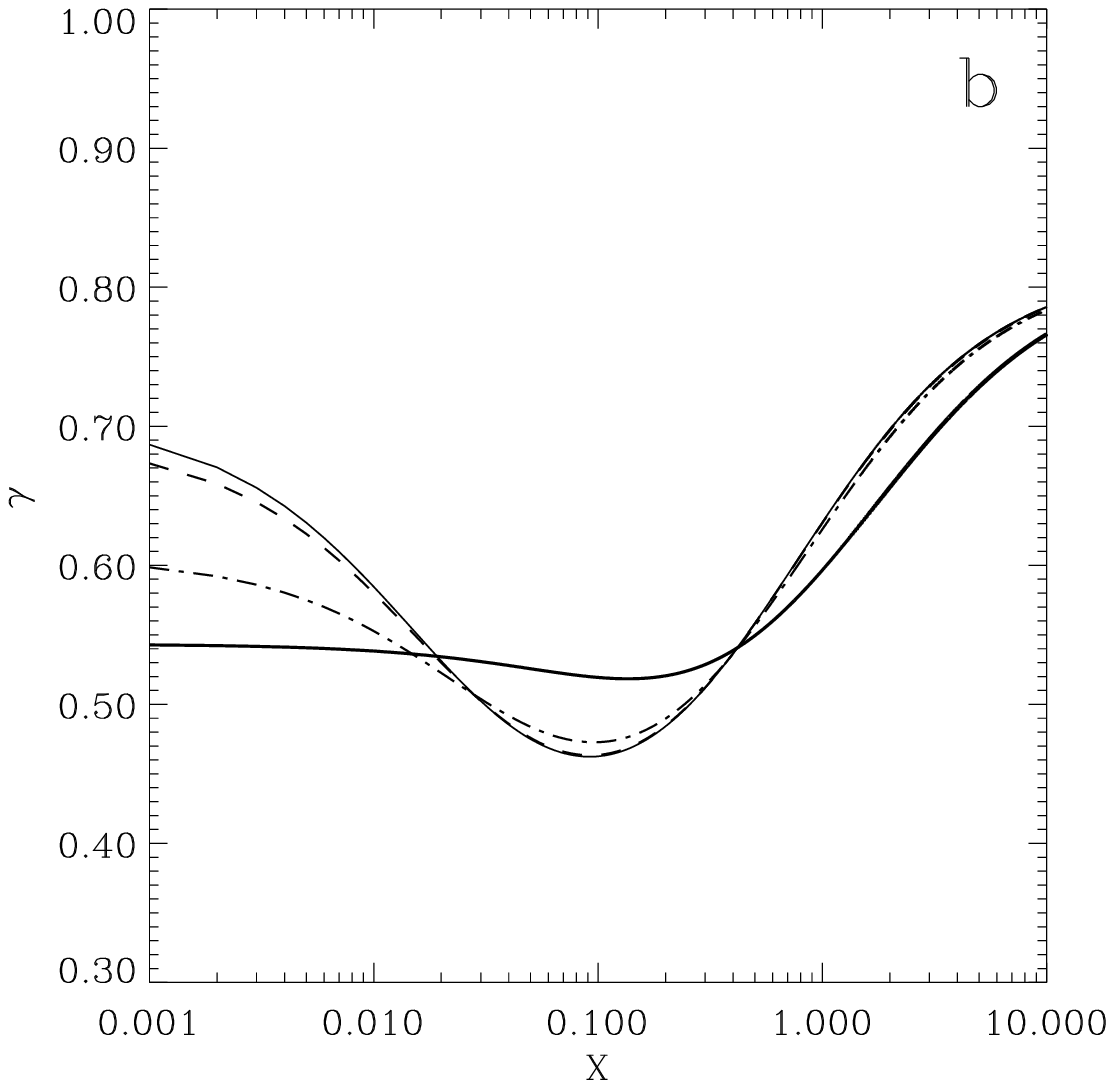}
\vspace{0.1cm}
\hspace{-0.01cm}\vspace{-0.1cm}
\epsfxsize=5cm 
\epsfbox{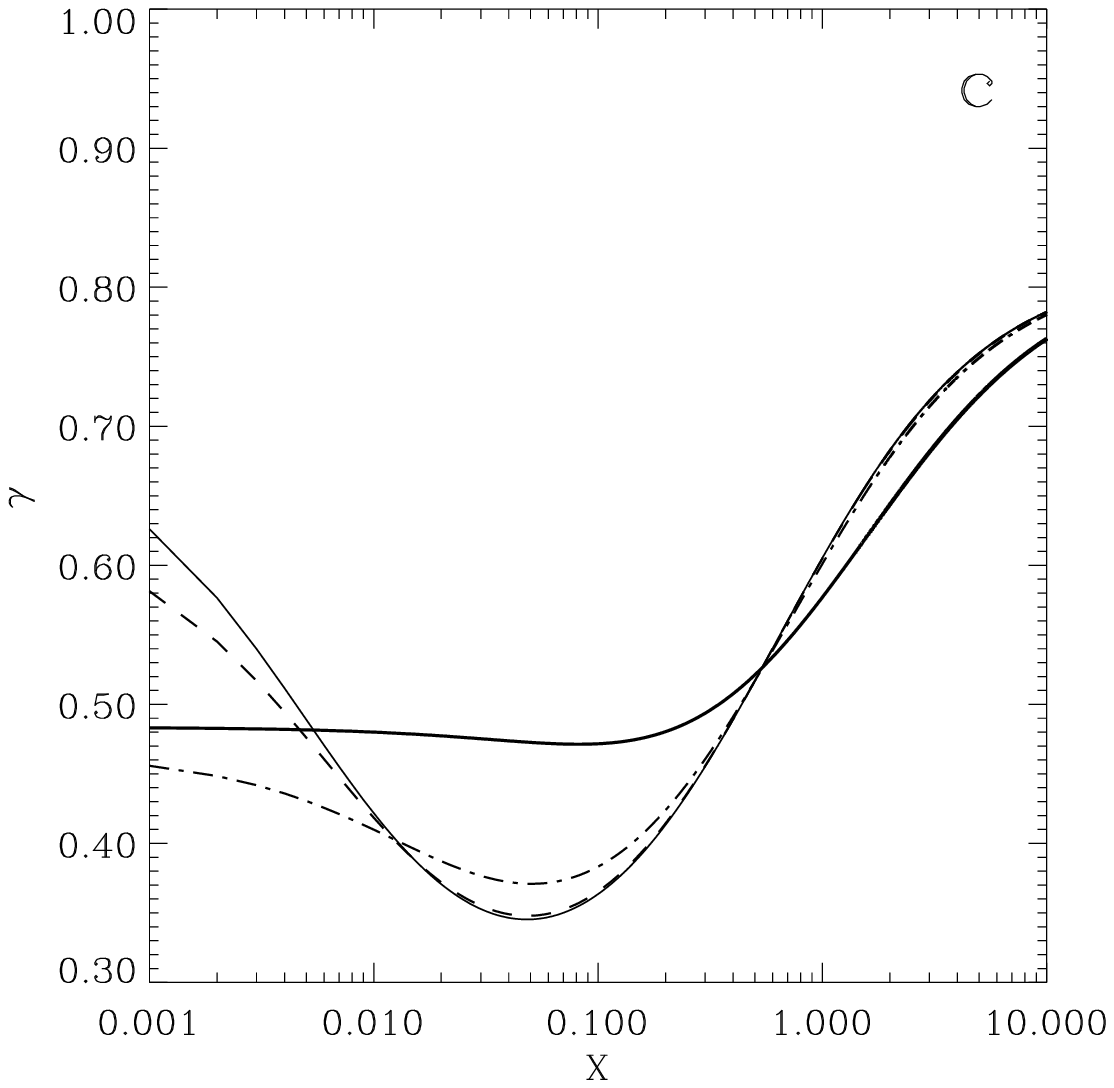}
\hspace{0.001cm} 
\epsfxsize=5cm 
\epsfbox{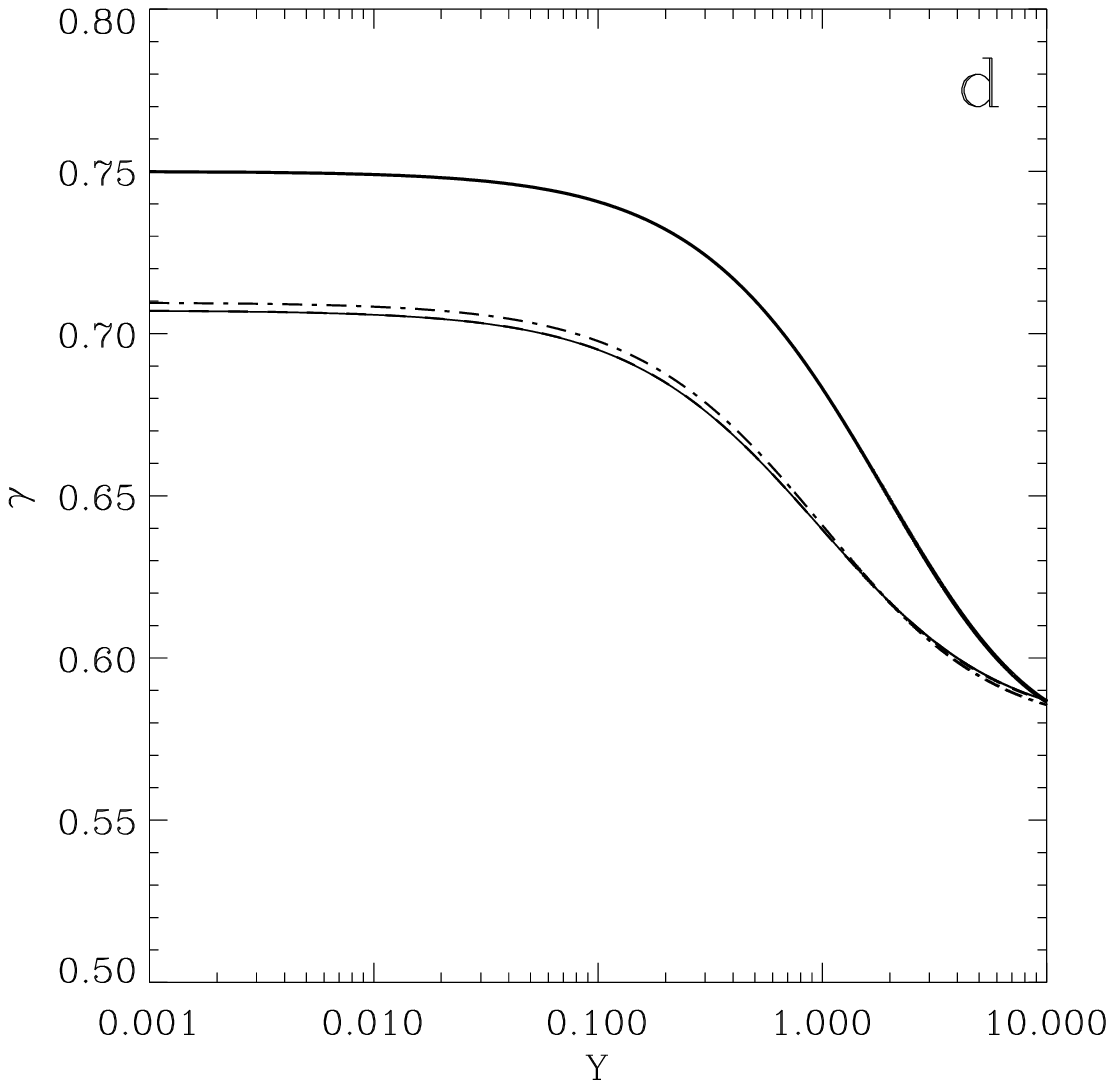}
\vspace{0.1cm}
\hspace{-0.01cm}\vspace{-0.1cm}
\epsfxsize=5cm 
\epsfbox{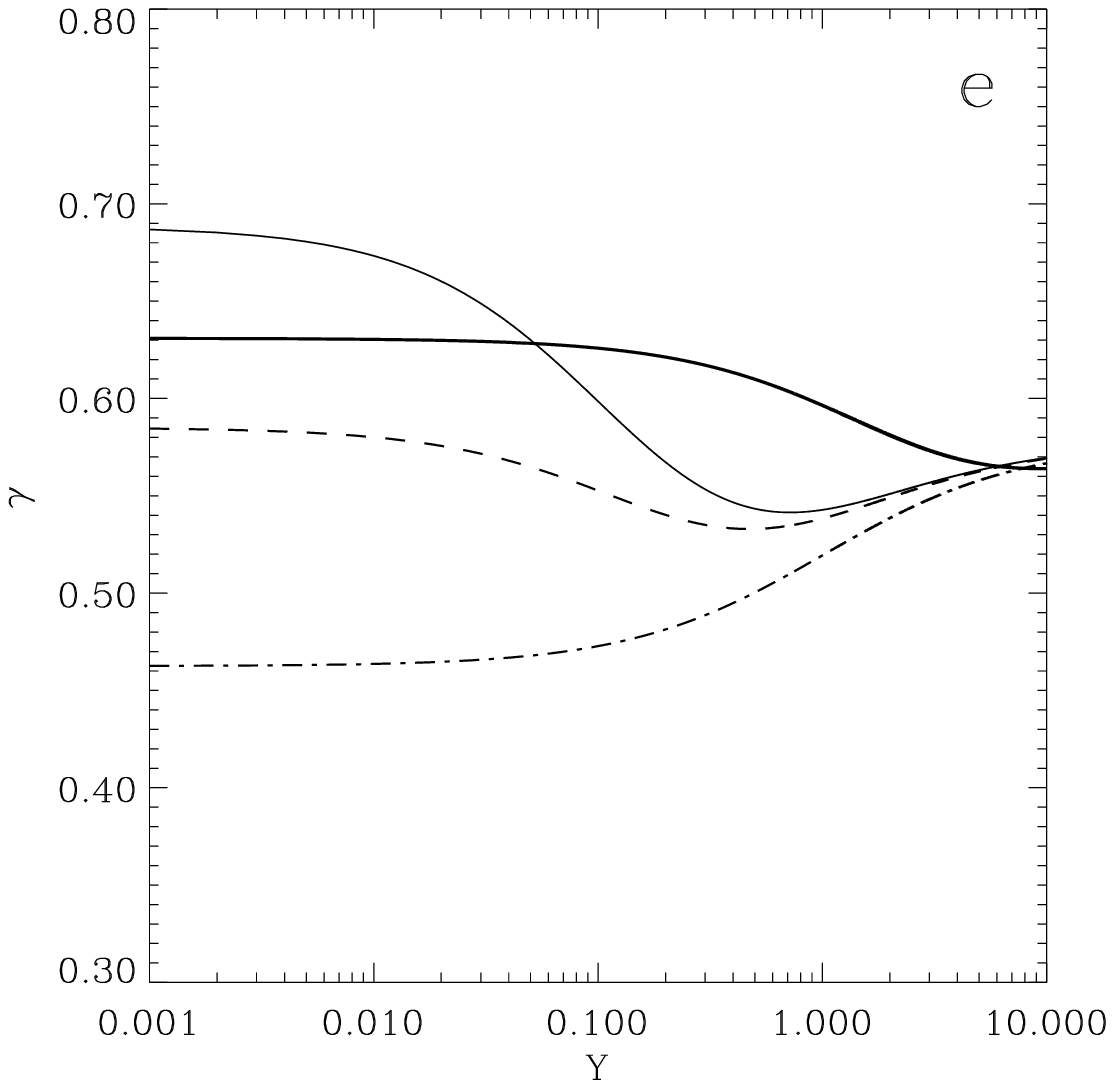}
\hspace{0.001cm} 
\epsfxsize=5cm 
\epsfbox{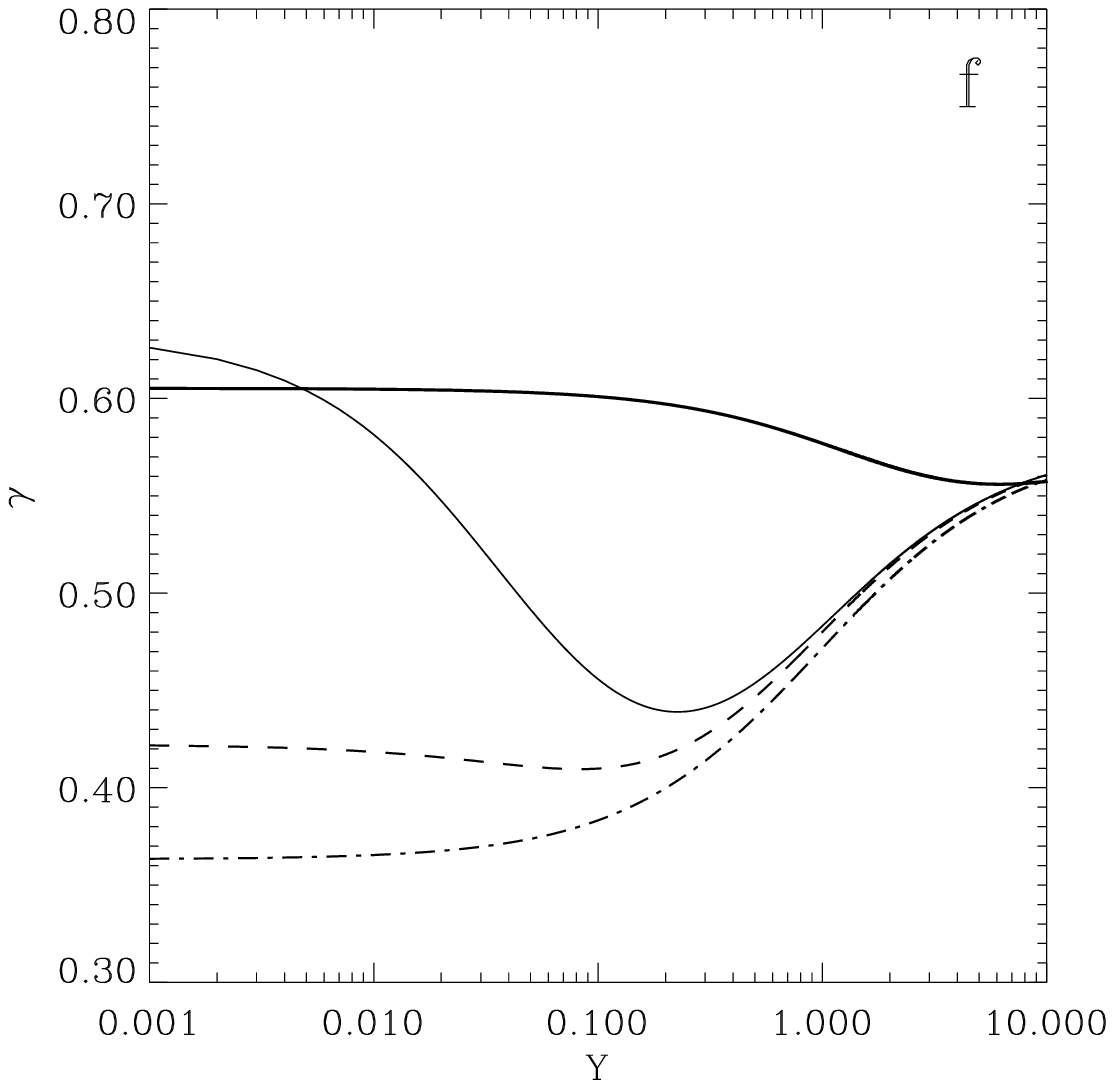}
\vspace{0.1cm}
    \caption{ {\it  
a) $\gamma(X,Y_n,\xi=1)$ at $Y_n=0.001$ (thin solid line), $Y_n=0.01$
(dashed line), $Y_n=0.1$ (dashed-dotted line) and $Y_n=1$ (thick solid 
line). b) and c) are the same as in a) but for $\xi=5$ and $\xi=10$ 
respectively. d) $\gamma(X_n,Y,\xi=1)$ for $X_n$=0.001 (thin solid 
line),$X_n$= 0.01  (dashed line), $X_n$=0.1 (dashed-dotted line); and 
$X_n$=1, (thick solid line); e) and f) are the same as in d) but for 
$\xi=5$ and $\xi=10$ respectively.}}
\end{figure}
Let us assume that the variance of AE signal is greater than those of the PS and 
CMB signals. In this case a typical value of parameter $\gamma(X,Y,\gth _S)$
is $\gamma\simeq \frac{1}{\sqrt{3}}$. If the angular resolution of the antenna 
beam corresponds to $\xi\simeq 1$ then after averaging the signal the  
parameter $\gamma(X=0,Y,\gth _S)$ monotonically transforms from 
$\gamma\simeq \frac{1}{\sqrt{3}}$ to $\gamma\simeq \frac{1}{\sqrt{2}}$
 (see Fig.6). This means that the rate of the maxima clusterisation 
decreases monotonically. However, we would like to point out that for $\xi>1$, 
for example $\xi\simeq 10$, which is close to the RATAN-600 antenna 
characteristics, the situation changes drastically. In this case  during 
the averaging procedure the value of the $\gamma$ parameter decreases with time 
while the rate of clusterisation increases. This process continues up to the 
moment when $\gamma$ reaches its minimum at some moment of time  $t_{cr}$.
At $t>t_{cr}$ the value of $\gamma$ goes up and the rate of clusterisation
monotonically  decreases.  We name  the point $t=t_{cr}$ when $\gamma$ 
reaches its minimum, as ``the turning point'' of the maximum clusterisation. 

As one can see from Fig. 6 the existence of such points is natural both 
for AE+CMB signals and for all types of foregrounds. For a specific value
 of $\xi$ (for a specific antenna) the turning points form a line $X=X(Y)$ in the 
plane$(X,Y)$. This non-trivial behavior of the parameter $\gamma$ gives new 
possibilities for separating the primordial CMB signal from noise.

\begin{table}
\caption{Numerical values of $\gamma$-parameter for different cosmological 
models.}
\vspace{0.3cm}
\centering
\begin{tabular}{|ccccccccccc|l|} \hline 
  &  &    & & &  SCDM  &  &  &           & & & $\gL CDM$\\ \hline
  &  & $H_0=50$ &     &                  & $\|$ &  &  & $H_0=75$  & & & $H_0=75$ \\ \hline
  &  & $\gW_b$ &  &  & $\|$ &  &  & $\gW_b$  &   &    & $\gW_b$ \\ \hline
0.0125 & $\|$ & 0.025 & $\|$ & 0.05 & $\|$ & 0.0125 & $\|$ & 0.025 & $\|$ & 0.05& 0.0125 \\ \hline
  &  & $\gamma$ &  &  & $\|$ &  &  & $\gamma$  &   &    & $\gamma$ \\ \hline
0.82 & $\|$ & 0.77 & $\|$ & 0.78 & $\|$ & 0.82 & $\|$ & 0.76 & $\|$ & 0.79& 0.72 \\ \hline
\end{tabular}
\end{table}
\begin{figure}
\vspace{-0.01cm}\hspace{2cm}
\epsfxsize=8cm 
\epsfbox{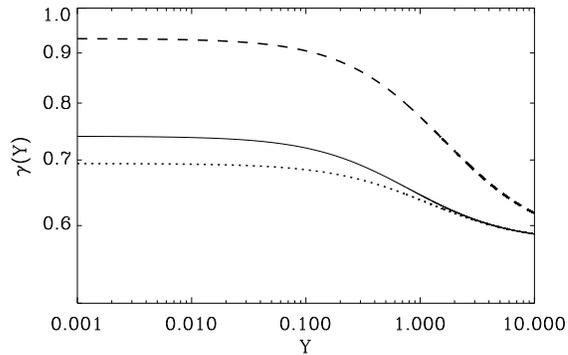}
     \caption{{\it Parameter $\gamma_{tot}(Y)$ for $l_S=10^2$ (dashed line),
$l_S=5\times 10^2$ (solid line), and $l_S=10^3$ (dotted line).}}
\end{figure}

Finally we consider in more detail the dependence 
of $\gamma$ on $Y$ when $\gth_S\le R_d$ which corresponds to the range of the 
spectrum with $l\le l_d\approx 10^3$. For this range of $l$ we can neglect 
the damping decrease of the CMB spectrum. At $10^2<l<10^3$ CMB spectrum has 
a few maxima and we will use the exact numerical description of the spectrum
 of CMB.
For simplicity we will consider a case where PS noise is not important for 
this range of $l$ and we will consider CMB+AE signal only. We have plotted the 
corresponding dependence $\gamma_{tot}(Y,\gth_S)$ in Fig. 7 for a SCDM power 
spectrum and AE noise and for $l_S=10^2$, $l_S=5\times 10^2$ and $l_S=10^3$,
where $l_S\sim \gth_S^{-1}$.

 As one can see from this 
figure the shape of  $\gamma_{tot}(Y,\gth_S)$ is monotonic and increases from 
$\gamma_{tot}\simeq \frac{1}{\sqrt{3}}$ up to $\gamma_{tot}\simeq0.92$ 
(for $\gth_S=10^{-2}$) if $Y$ decreases. This means that in the range 
$10^2<l_S<10^3$ we have practically the same behavior of $\gamma$ parameter 
as in Figure 6d.

\section{Conclusions}
The main goal of this paper was to develop the cluster analysis of the 
one-dimensional random fields in the application to the RATAN-600 observational 
scans for measurements of the CMB anisotropy. We analyzed the properties of the 
peak clusterisation in the RATAN-600 scans and demonstrated how the high angular
resolution of the antenna of RATAN-600 can be used to reveal noise related 
to atmospheric emission and unresolved point sources in the observational 
data.

We believe that the cluster analysis method will find its place in the future data 
reduction of ground based experiments together with other published methods
of AE filtration.

\vspace{0.1cm}
\noi
{\bf Acknowledgments}

\noi 
 P.N. is grateful to the staff of TAC, and NORDITA for providing
excellent working conditions during his visit to these institutions. 
This investigation was supported in part by a grant ISF MEZ 300,  
 by a grant INTAS 97-1192, by the Danish Natural Science 
Research Council through grant No 9701841 and  by Danmarks 
Grundforskningsfond through its support for the 
establishment of the Theoretical Astrophysics Center.

\vspace{0.2cm}
\noi
{\bf References}

\noi
1. B. Melchiorri, M. de Petris, G.D'andreta, G. Guarini, F. Melchiorri 
and M. Signore; {\it Astrophys. J.}, {\bf 471}, 52, (1996).

\noi
2. 
O. Lay, N. Halverson, Astroph/9905369 (1999).

\noi
3.  W., Hu,  http:/www/sns.ias.edu.

\noi
4. 
M. Tegmark, D. Eisenstein, W.Hu, A. de Oliviera-Costa; Astroph/9905257 (1999).

\noi
5.  P. Naselsky,  and D. Novikov, {\it Astrophys. J.}, {\bf 444}, L1, (1995).

\noi
6. D. Novikov and H. J\o rgensen, {\it Astrophys. J.}, {\bf 471}, 521, (1996).

\noi
7.  Bond., Y.R., and G.Efstathiou, {\it Mon. Not. Roy. Astro. Soc}, 
{\bf 226}, 655, (1987).

\noi
8.  I.S. Gradstein, \@ Ryzhik, I.M. "{\it Tables of Integrals, 
Series and Products} Ed. by Alan Jeffrey Academic Press, (1980).

\noi
9.  P. Naselsky,  I. Novikov, Yu Parijskij, P. Tcibulev; Internationa 
Journal of Moden Physics, {\bf 8}, n5, (1999)

\noi
10. G. Jungman, M. Kamionkowski, A. Kosowsky, D. Spergel; {\it Phys. Rev. D}
{\bf 54}, 1332, (1996).

\noi
11. \noi
A.V. Chepurnov; PhD thesis (1997).

\end{document}